\documentclass[pre,floatfix,showpacs,twocolumn]{revtex4}

\usepackage{epsfig}
\usepackage{amssymb}

\begin{document}

\title{Phase Field Modeling of Fracture and Stress Induced Phase Transitions}
\author{R. Spatschek}
\author{C. M\"uller-Gugenberger}
\author{E. Brener}
\affiliation{Institut f\"ur Festk\"orperforschung, Forschungszentrum J\"ulich, D-52425 J\"ulich, Germany}
\author{B. Nestler}
\affiliation{Institute of Computational Engineering, University of Applied Sciences D-76133 Karlsruhe, Germany}

\date{\today}

\begin{abstract}
We present a continuum theory to describe elastically induced phase transitions between coherent solid phases.
In the limit of vanishing elastic constants in one of the phases, the model can be used to describe fracture on the basis of the late stage of the Asaro-Tiller-Grinfeld instability.
Starting from a sharp interface formulation we derive the elastic equations and the dissipative interface kinetics.
We develop a phase field model to simulate these processes numerically;
in the sharp interface limit, it reproduces the desired equations of motion and boundary conditions.
We perform large scale simulations of fracture processes to eliminate finite-size effects and compare the results to a recently developed sharp interface method.
Details of the numerical simulations are explained, and the generalization to multiphase simulations is presented.
\end{abstract}


\maketitle

%%%%%%%%%%%%%%%%%%%%%%%%%%%%%%%%%%%%%%%%%%%%%%%%%%%%%%%%%%%%%%%%%%%%%
\section{Introduction}

The propagation of cracks is very important for many applications and a central topic in physics and materials science.
The most fundamental basis of understanding fracture traces back to Griffith \cite{Griffith21};
according to his findings, the growth of cracks is determined by a competition of a release of elastic energy and a simultaneous increase of the surface energy if a crack extends.
Although much progress has been made in understanding the striking features of cracks \cite{Freund98}, the mechanisms which determine the dynamics of crack propagation are still under heavy debate.
A typical description of cracks starts on the atomic level and interprets the propagation by successive breaking of bonds;
it is obvious that the theoretical predictions significantly depend on the
underlying empirical models of the atomic properties (see for example
\cite{Hauch99}).
A rather complementary approach takes into account effects like plasticity, which can lead to extended crack tips (finite tip radius $r_0$) \cite{Langer00}.
Recent experimental investigations of fracture in brittle gels \cite{Fineberg} possibly reveal macroscopic scales.
It is obvious that under these circumstances a full modeling of cracks should not only determine the crack speed but also the entire crack shape and scale self-consistently.

During the past years, phase field modeling has emerged as a promising approach to analyze fracture by continuum methods.
Recent phase field models go beyond the microscopic limit of discrete models, and encompass much of the expected behavior of cracks \cite{Karma,Henry04,Aranson00,Eastgate02};
However, a significant feature of these descriptions is that the scale of the growing patterns is always set by the phase field interface width, which is a purely numerical parameter and not directly connected to physical properties;
therefore these models do not possess a valid sharp interface limit.
Alternative descriptions, which are intended to investigate the influence of elastic stresses on the morphological deformation of surfaces due to phase transition processes, are based on macroscopic equations of motion.
But they suffer from inherent finite time singularities which do not allow steady state crack growth unless the tip radius is again limited by the phase field interface width \cite{Kassner01}.
Very different approaches which are not based on a phase field as order parameter introduce a tip scale selection by the introduction of complicated nonlinear terms in the elastic energy for high strains in the tip region \cite{Marconi05}, requiring additional parameters.

Recently, we developed a minimum theory of fracture \cite{Brener03} which is only based on well-established thermodynamical concepts.
This is also motivated by experimental results showing that many features of crack growth are rather generic \cite{Fineberg99};
among them is the saturation of the steady state velocity appreciably below the Rayleigh speed and a tip splitting for high applied tension.
This theory describes crack growth as a consequence of the Asaro-Tiller-Grinfeld (ATG) instability \cite{ATG} in the framework of a continuum theory.
Mass transport at the extended crack tips can be either due to surface diffusion or a phase transformation process.
The latter has been investigated numerically by phase field simulations \cite{Spatschek06} and sharp interface methods \cite{Pilipenko06}.
It turned out that the phase field simulations were still significantly influenced by finite size effects and insufficient separation of the appearing length scales, and therefore the results did not coincide.
One central aim of the current paper is therefore to carefully extrapolate new phase field results obtained by large-scale computations.
As we will show, we then get a very convincing agreement of the approaches.
Also, we explain details of the phase field method and the underlying sharp interface equations in more detail.

The paper is organized as follows:
First, in Section \ref{SharpInterfaceSection}, we introduce the sharp interface equations to describe crack propagation as a phase transformation process.
The basic selection mechanisms for crack growth are reviewed in Section \ref{SelectionSection}.
We introduce a phase field description to solve the arising moving boundary problem in Section \ref{PhasefieldSection}.
We demonstrate the numerical separation of lengthscales and obtain results which are in excellent agreement with sharp-interface predictions (Section \ref{extrapolation}).
This part contains important new results concerning the underlying continuum theory of fracture.
In Section \ref{multiphase} we briefly explain how the model can be extended to systems consisting of multiple phases.
Detailed derivations of the sharp interface equations are given in Appendix \ref{SharpInterfaceAppendix}.
Since coherency of two solid phases leads to an unexpected expression for the chemical potential, its relevance is analyzed specifically for the motion of a planar interface (Appendix \ref{planarAppendix}).
In Appendix \ref{CoherentPhasefieldAppendix} it is demonstrated that the presented phase field model recovers this effect in the sharp interface limit.
Finally, details of the numerical implementation of the phase field model are given in Appendix \ref{ImplementationAppendix}.

%%%%%%%%%%%%%%%%%%%%%%%%%%%%%%%%%%%%%%%%%%%%%%%%%%%%%%%%%%%%%%%%%%%%%
\section{Sharp interface description}
\label{SharpInterfaceSection}

The fracture process in \cite{Spatschek06} is interpreted as a first order phase transition from the solid to a ``dense gas phase'', driven by elastic effects.
More generally, we investigate the transition between two different solid phases.
In the limiting case that one phase is infinitely soft, crack propagation can be studied.
A central simplification is due to the assumption of equal mass density $\rho$ in both phases and the condition of coherency at the interface, i.e.~the displacement field $u_i$ is continuous across the phase boundary,
\begin{equation} \label{coherency}
u_i^{(1)}=u_i^{(2)},
\end{equation}
where the upper indices indicate the different phases.
The strain $\epsilon_{ik}$ is related to the displacement field $u_i$ by
\begin{equation}
\epsilon_{ik}^{(\alpha)}= \frac{1}{2}\left( \frac{\partial u_i^{(\alpha)}}{\partial x_k} + \frac{\partial u_k^{(\alpha)}}{\partial x_i} \right).
\end{equation}
Strain and stress $\sigma_{ik}$ are connected through Hooke's law;
for the specific case of isotropic materials, it reads
\begin{equation}
\sigma_{ik}^{(\alpha)} = \frac{E^{(\alpha)}}{1+\nu^{(\alpha)}} \left(\epsilon_{ik}^{(\alpha)} + \frac{\nu^{(\alpha)}}{1-2\nu^{(\alpha)}} \delta_{ik}\epsilon_{ll}^{(\alpha)} \right)
\end{equation}
for each phase.
Here, $E$ and $\nu$ are elastic modulus and Poisson ratio, respectively.
We note that eigenstrain contributions due to different unit cells of the phases are not considered here for brevity, but they can easily be introduced \cite{Brener06}.
For simplicity, we assume a two-dimensional plane-strain situation.

\begin{figure}
\begin{center}
\epsfig{file=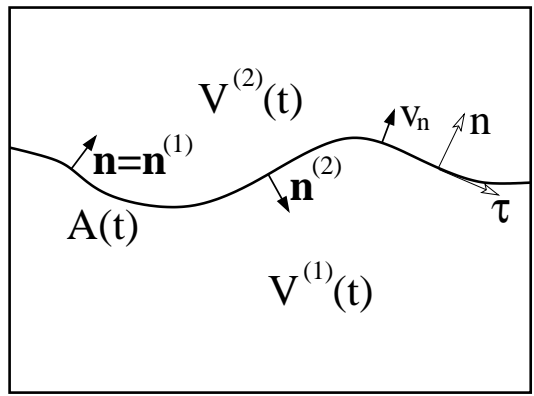, width=7cm}
\caption{Geometry of the phase transition scenario. Phase transitions between phases 1 and 2 are possible and lead to interface motion with local normal velocity $v_n$. The volumes of the two phases and the interface $A(t)$ are therefore time-dependent.}
\label{fig1}
\end{center}
\end{figure}

All following relations are obtained in a consistent way from variational principles, and this is described in detail in Appendix \ref{SharpInterfaceAppendix}.
The equations of dynamical elasticity are
\begin{equation} \label{dynelast}
\frac{\partial \sigma_{ik}^{(\alpha)}}{\partial x_k} = \rho \ddot{u}_i,
\end{equation}
for each phase.
On the interface, we obtain the expected continuity of normal and shear stresses
\begin{equation} \label{elbc}
\sigma_{in}^{(1)} = \sigma_{in}^{(2)}.
\end{equation}
Here, the index $n$ denotes the normal direction of the interface, with the perpendicular tangential direction $\tau$ (see Fig.~\ref{fig1}).
We mention that this equation holds only for the specific conditions of equal mass densities and coherency at the interface.
In other cases one obtains more general relations for momentum conservation at the interface, which also involve the interface normal velocity $v_n$ \cite{Freund98}.

The elastic contribution to the chemical potential at the interface for each phase is
\begin{equation} \label{chempot}
\mu^{(\alpha)}_{el} = \Omega \left(\frac{1}{2} \sigma_{\tau\tau}^{(\alpha)}\epsilon_{\tau\tau}^{(\alpha)} - \frac{1}{2} \sigma_{nn}^{(\alpha)}\epsilon_{nn}^{(\alpha)} - \sigma_{n\tau}^{(\alpha)}\epsilon_{n\tau}^{(\alpha)} \right).
\end{equation}
Since the chemical potential has the dimension energy per particle, we introduced the atomic volume $\Omega$.
It is quite remarkable that the normal and shear contributions enter into the expression with negative sign, in contrast to the natural expectation $\tilde{\mu}_{el}=\Omega\sigma_{ik}\epsilon_{ik}/2$, which is the potential energy density.
The reason for this modification is the coherency constraint which has to be fulfilled at the interface.
An illustrative example to understand this unexpected expression for the chemical potential is given in Appendix \ref{planarAppendix}.
However, this effect is only important for solid-solid transformations.
For crack propagation, where we assume that the new phase inside the crack is infinitely soft, the normal and shear stresses vanish at the interface, and therefore the discrepancy between the chemical potential (\ref{chempot}) and the naive guess $\tilde{\mu}_{el}$ disappears.

We would also like to mention that the equality of the mass density and the coherency leads to the absence of kinetic energy contributions to the chemical potential.
The reason is that such a contribution is continuous across the interface;
finally, only the chemical potential difference $\mu^{(1)}-\mu^{(2)}$ enters into the equation of motion for the interface, and therefore such a term would cancel.
Nevertheless, we note that the kinetic energy contribution would enter into the chemical potential with negative sign, i.e. $-\rho \dot{u}_i^2$, because in the Lagrangian the kinetic and potential energy contributions appear with opposite sign.
Kinetic contributions may play a role if instead of a phase transformation process, surface diffusion along a free boundary drives the evolution.

For the motion of the interface, surface energy is also taken into account.
Since it does not couple to the elastic terms, it simply gives an additional contribution to the chemical potential,
\[
\mu_s = \Omega\gamma\kappa,
\]
with the local interface curvature $\kappa$ and the surface energy density $\gamma$.
The curvature is positive if phase 1 is convex.
Then the motion of the interface due to a phase transition process is described by the local normal velocity ($D$ is a kinetic coefficient with dimension $[D]=\mathrm{m^2/s}$)
\begin{equation} \label{normalvel}
v_n = -\frac{D}{\gamma \Omega} (\mu^{(1)}_{el} - \mu^{(2)}_{el} + \gamma\kappa),
\end{equation}
which is positive if phase 1 grows.
The set of equations (\ref{coherency})-(\ref{normalvel}) describes the dynamics of the system.
We point out that it leads to a complicated free boundary problem, and the arising interfacial patterns are self-consistently selected during this nonequilibrium process if external forces are applied to the system.
For an initially almost flat interface between a soft and a hard phase, which is nonhydrostatically strained, first the ATG instability develops:
Long wave morphological perturbations lead to a decrease of the total energy and the formation of deep notches, similar to cracks (see e.g.~\cite{Kassner01,Yang}).
As we have shown in \cite{Brener03,Spatschek06,Pilipenko06}, it is essential to include the inertial contributions, because otherwise a steady state growth of these cracks is impossible, and the system collapses into the finite time cusp singularity of the ATG instability.

%%%%%%%%%%%%%%%%%%%%%%%%%%%%%%%%%%%%%%%%%%%%%%%%%%%%%%%%%%%%%%%%%%%%%
\section{Crack propagation: Selection principles}
\label{SelectionSection}

In the case that one phase is infinitely soft, crack propagation can be studied.
Here, growth of the crack is based on a phase transition of the solid matrix to a ``dense gas phase'' which has the same density as the solid.
In this sense, it is similar to other models of fracture based on a non-conserved order parameter \cite{Karma,Henry04,Aranson00}.
The crucial difference is that the current model is based on well-defined sharp interface equations, and therefore the predictions do not depend on inherently numerical parameters like a phase field interface width.
However, numerically, it requires a tedious separation of scales to obtain these results;
this is described in the next section.

Understanding fracture as a phase transition process offers many numerical advantages, as phase field models can be derived to solve this moving-boundary problem.
We point out that the underlying selection principles which allow a steady state crack growth with propagation velocities well below the Rayleigh speed, tip blunting and branching for high driving forces are rather generic and are similarly valid for models with conserved order parameters.
In \cite{Brener03}, we derived the similar equations of motion if instead of a phase transition, mass transport is due to surface diffusion along the free crack boundary.

In the latter case, the elastic boundary conditions are replaced by
\begin{equation}
\sigma_{in}^{(sd)}= - \rho\dot{u}_i v_n,
\end{equation}
and the chemical potential becomes
\begin{equation} \label{sdchempot}
\mu^{(sd)} = \Omega \left( \frac{1}{2}\sigma_{ik}\epsilon_{ik} - \frac{1}{2}\rho \dot{u}_i^2+\gamma\kappa \right).
\end{equation}
It differs from the expression (\ref{chempot}) first by the elastic energy density, because no coherency constraints have to be fulfilled here.
Second, the kinetic energy density appears here, because it does not cancel in the derivation from the Lagrangian.
The equation of motion for the interface is replaced by
\begin{equation} \label{sdnormalvel}
v_n^{(sd)}= \frac{D^{(sd)}}{\gamma\Omega} \frac{\partial^2\mu^{(sd)}}{\partial \tau^2}
\end{equation}
with the surface diffusion coefficient $D^{(sd)}$.

In both cases, stresses on the boundary of the crack tip with finite radius $r_0$ scale as
\begin{equation}
\sigma\sim K r_0^{-1/2},
\end{equation}
and the curvature behaves as $\kappa\sim1/r_0$.
Therefore, all contributions to the chemical potentials scale like $\mu\sim r_0^{-1/2}$, and this is ultimately the reason for the cusp singularity of the Grinfeld instability and the impossibility of a steady-state crack growth, if only static elasticity is taken into account:
Then, the equations of motion (\ref{normalvel}) or (\ref{sdnormalvel}) can be rescaled to an arbitrary tip radius which is not selected by the dynamical process.
The explanation is that the linear theory of elasticity and surface energy define only one lengthscale, the Griffith length, which is macroscopic, but do not provide a microscopic scale which allows the selection of a tip scale.
Formally, the equations of motion depend only on the dimensionless combinations $vr_0/D$ for the phase transition dynamics and $vr_0^3/D^{(sd)}$ for surface diffusion;
the radius $r_0$ and the steady state velocity $v$ therefore cannot be selected separately;
any rescaling which maintains the value of the product would therefore describe another solution.
The situation changes if inertial effects are taken into account, which is reasonable for fast crack propagation.
Then additionally the ratio $v/v_R$ ($v_R$ is the Rayleigh speed) appears in the equations of motion, and therefore a rescaling is no longer possible.
Instead, $D/v_R$ for the phase transition dynamics and $(D^{(sd)}/v_R)^{1/3}$ for surface diffusion set the tip scale.
Thus, we conclude that fast steady state growth of cracks is possible if inertial effects are taken into account.
More formal analyses also including rigorous selection mechanisms due to the suppression of growing crack openings far behind the tip are given in \cite{Spatschek06} for phase transition processes and in \cite{Brener03} for surface diffusion.

This analysis shows that the selection principles which allow a fast steady state growth of cracks are similar for the simple phase transition process studied here and surface diffusion.
The latter does not require the introduction of a dense gas phase inside the crack and obeys conservation of the solid mass itself.
Even more, surface diffusion can be understood in a generalized sense as plastic flow in a thin region around the extended tip which can be effectively described in the spirit of a lubrication approximation.
Therefore, many general statements obtained for the phase transition dynamics can also be used for crack growth propelled by surface diffusion.
The latter is more tedious to implement numerically, since the equation of motion (\ref{sdnormalvel}) is of higher order \cite{voigt}.

For both mechanisms, a tip splitting is possible for high applied tensions due to a secondary ATG instability:
Since $\sigma\sim Kr_0^{-1/2}$ in the tip region and the local ATG length is $L_G\sim E\gamma/\sigma^2$, an instability can occur, provided that the tip radius becomes of the order of the ATG length.
In dimensionless units, this leads to the prediction $\Delta_{split}\sim {\cal O}(1)$.

%%%%%%%%%%%%%%%%%%%%%%%%%%%%%%%%%%%%%%%%%%%%%%%%%%%%%%%%%%%%%%%%%%%%%
\section{Phase field model}
\label{PhasefieldSection}

To describe systems with moving boundaries according to the equations of motion developed above, we implemented a phase field model.
Let $\phi$ denote the phase field with values $\phi=1$ for phase 1 and $\phi=0$ for phase 2.
The energy density contributions are
\begin{equation}
f_{el}=\mu(\phi)\epsilon_{ij}^2+\lambda(\phi)(\epsilon_{ii})^2/2
\end{equation}
for the elastic energy, with the interpolated shear modulus and Lam\'e coefficient
\begin{eqnarray}
\mu(\phi) &=& h(\phi)\mu^{(1)} + (1-h(\phi))\mu^{(2)}, \label{muinterpolated} \\
\lambda(\phi) &=& h(\phi)\lambda^{(1)} + (1-h(\phi))\lambda^{(2)} \label{lambdainterpolated}
\end{eqnarray}
where
\begin{equation}
h(\phi)=\phi^2(3-2\phi)
\end{equation}
interpolates between the phases, and the superscripts denote the bulk values.
The surface energy is
\begin{equation}
f_s(\phi)=3\gamma \xi (\nabla\phi)^2/2
\end{equation}
with the interface width $\xi$.
Finally,
\begin{equation}
f_{dw}=6\gamma\phi^2(1-\phi)^2/\xi
\end{equation}
is the well-known double well potential.
Thus the total free energy is given by
\begin{equation}
F = \int dV \left( f_{el}+f_s+f_{dw}\right).
\end{equation}
The elastodynamic equations are derived from the free energy by variation with
respect to the displacements $u_i$,
\begin{equation} \label{phase:eq2}
\rho \ddot{u}_i=-\left( \frac{\delta F}{\delta u_i} \right)_{\phi=const},
\end{equation}
and the dissipative phase fields dynamics follows from
\begin{equation} \label{phase:eq1}
\frac{\partial\phi}{\partial t} = -\frac{D}{3\gamma\xi} \left( \frac{\delta F}{\delta\phi} \right)_{u_i=const}.
\end{equation}
It has been shown in \cite{Kassner01} that in the quasistatic case, the above equations lead to the sharp interface equations (\ref{dynelast})-(\ref{normalvel}) if the interface width $\xi$ is significantly smaller than all physical lengthscales present in the system.
In Appendix \ref{CoherentPhasefieldAppendix}, it is illustrated that this model also correctly incorporates the modification of the chemical potential (\ref{chempot}) due to the coherency constraint.
Details of the numerical implementation are given in Appendix \ref{ImplementationAppendix}.

%%%%%%%%%%%%%%%%%%%%%%%%%%%%%%%%%%%%%%%%%%%%%%%%%%%%%%%%%%%%%%%%%%%%%
\section{Phase field modeling of crack propagation}
\label{extrapolation}

The central prediction of this theory of fracture is that a well-defined steady-state growth with finite tip radius and velocities appreciably below the Rayleigh speed is possible.
This also cures the problem of the finite-time cusp singularity of the Grinfeld instability.
These predictions have been confirmed by phase field simulations \cite{Spatschek06} and sharp interface methods \cite{Pilipenko06} which are based on a multipole expansion of the elastodynamic fields.
Surprisingly, it turned out that the obtained results seem to differ significantly:
In particular, the sharp interface method predicts a range of driving forces inside which the velocity of the crack is a monotonically decreasing function.
Here, we demonstrate that the discrepancy of results is due to finite size effects of the previous phase field results \cite{Spatschek06}, and that by careful extrapolation of large-scale simulations, a coinciding behavior is obtained.

We investigate crack growth in a strip geometry with fixed displacements at the upper and lower grip.
The multipole expansion technique \cite{Pilipenko06} is designed to model a perfect separation of the crack tip scale $D/v_R$ to the strip width $L$:
In most real cases, crack tips are very tiny, and therefore it is theoretically desirable to describe this limit.
For the phase field method, however, a finite strip width $L$ is necessary, and a good separation of the scales therefore requires time-consuming large-scale calculations.
We typically use strip lengths $2L$ and shift the system such that the tip remains in the horizontal center.
This allows to study the propagation for long times until the crack reaches a steady state situation.
Apart from this finite size restriction, we had to introduce the interface width $\xi$ as a numerical parameter, and the phase field method delivers quantitative results only in the limit that all physical scales are much larger than this lengthscale.
The latter has to be noticeably larger than the numerical lattice parameter $\Delta x$, but the results show that the choice $\xi=5\Delta x$ is sufficient.
We therefore have to satisfy the hierarchy relation
\begin{equation} \label{ScaleSeparation}
\xi \ll \frac{D}{v_R} \ll L,
\end{equation}
which is numerically hard to achieve.
We developed a parallel version of the phase field code which is run on up to 2048 processors, with system sizes up to $8192\times 4096\cdot(\Delta x)^2$.
All computations are performed on the supercomputers JUMP and JUBL operated at the Research Center J\"ulich.

For the strip geometry, a dimensionless driving force $\Delta$ is defined as
\begin{equation}
\Delta = \frac{\delta^2 (\lambda + 2\mu)}{4\gamma L},
\end{equation}
with $\delta$ being a fixed displacement by which the strip is elongated vertically.
The elastic constants of the new phase inside the crack are zero.
The value $\Delta=1$ corresponds to the Griffith point.
All calculations are done with Poisson ratio $\nu=1/3$.

\begin{figure}
\begin{center}
\epsfig{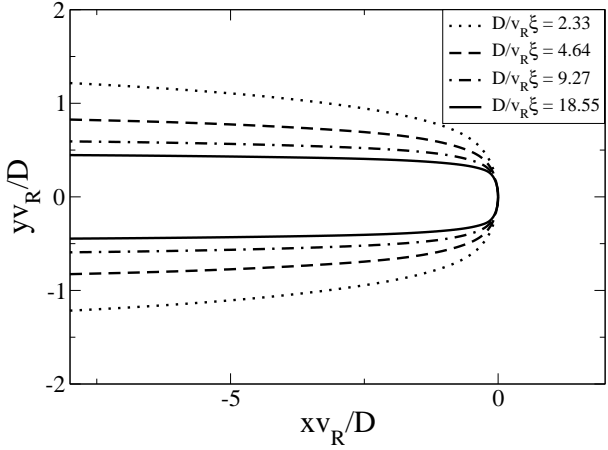}
\caption{Crack shapes for different scale separations $D/v_R\xi$ and fixed ratio $L v_R/D=11.03$; the aspect ratio of the system is $2:1$. The driving force is $\Delta=1.4$. By improvement of the separation, the crack opening is reduced, and finally the boundaries become straight parallel lines.}
\label{CrackOpenings}
\end{center}
\end{figure}

In \cite{Pilipenko06}, it was shown that close to the Griffith point, dissipation free solution exists in the framework of the model:
In this regime $1<\Delta<1.14$ an additional microscopic length scale is needed to select the small tip radius which is no longer determined by the ratio $D/v_R$.
This can here be mimicked by the phase field interface width and was already done in \cite{Spatschek06}.
Here, we focus on the more interesting regime of higher driving forces, but still below the threshold of instability.
Typical crack shapes in the vicinity of the tip are shown in Fig.~\ref{CrackOpenings}.

\begin{figure}
\begin{center}
\epsfig{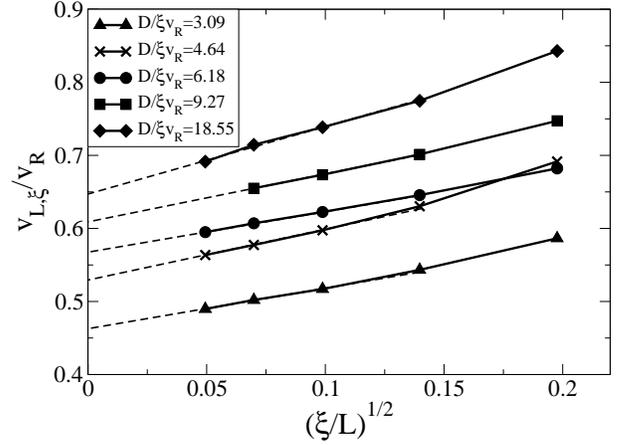}
\caption{First step of the extrapolation procedure for the dimensionless velocity $v/v_R$.
The system size $L/\xi$ is increased and the ratio $D/\xi v_R$ is kept fixed for each curve.
For each ratio, an extrapolated velocity $v_\xi/v_R$ corresponding to an infinite system size is obtained, as indicated by the dashed lines.
The driving force is $\Delta=1.4$.}
\label{FirstExtrapolation}
\end{center}
\end{figure}

To fulfill the scale separation (\ref{ScaleSeparation}), we perform a double extrapolation of the obtained steady state velocities $v_{L,\xi}$ (the subscripts indicate the additional non-resolved length scale dependencies).
In the first step, we extend the simulations to an infinite system size.
Therefore, we decrease the ratio $\xi/L\to 0$ for fixed tip scale ratio $D/\xi v_R$.
This step is demonstrated in Fig.~\ref{FirstExtrapolation} for $\Delta=1.4$.
Here, the dimensionless propagation velocity $\tilde{v}_{L,\xi}=v_{L,\xi}/v_R$ is plotted as function of the inverse square root of the system size $(\xi/L)^{1/2}$.
In this representation, the data for the larger systems can be extrapolated linearly to infinite system sizes, since we numerically get a scaling
\begin{equation}
\tilde{v}_{L,\xi}(\Delta, \frac{D}{v_R\xi}, \frac{L}{\xi}) = \tilde{v}_{\xi}(\Delta, \frac{D}{v_R\xi}) + \alpha \left( \frac{\xi}{L} \right)^{1/2}
\end{equation}
for large systems, $\xi/L\ll 1$, with a constant $\alpha>0$ for each curve.
Since the separation of $D/v_R$ to $\xi$ is still imperfect, the extrapolated values $\tilde{v}_{\xi}(\Delta, D/v_R\xi)$ do not yet cumulate to a single point, and therefore a second extrapolation step is necessary.

\begin{figure}
\begin{center}
\epsfig{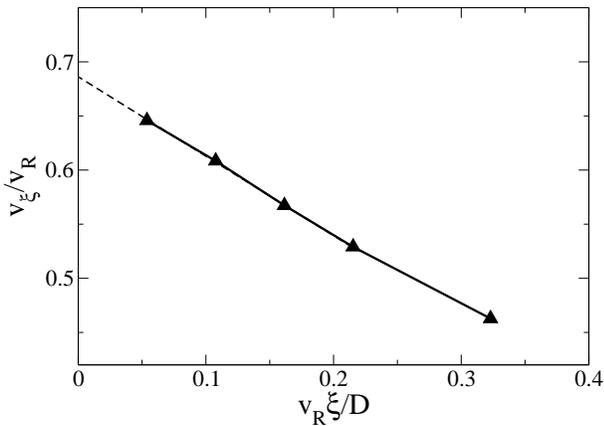}
\caption{Second extrapolation step to obtain the sharp interface velocity $v$. The extrapolated velocities obtained from Fig.~\ref{FirstExtrapolation} are plotted as function of the scale separation parameter $v_R/\xi D$. In this example $\Delta=1.4$ is used.}
\label{SecondExtrapolation}
\end{center}
\end{figure}

Hence, in Fig.~\ref{SecondExtrapolation}, the dependence of the velocity $v_{\xi}/v_R$ on the separation parameter $v_R\xi/D$ for $\Delta=1.4$ is shown.
The extrapolated values from Fig.~\ref{FirstExtrapolation} are used, and we obtain a scaling
\begin{equation}
\tilde{v}_{\xi}(\Delta, \frac{D}{v_R\xi}) = \tilde{v}(\Delta) - \beta \frac{v_R\xi}{D}
\end{equation}
with a constant $\beta>0$ and the dimensionless sharp interface limit velocity $\tilde{v}=v/v_R$.

\begin{figure}
\begin{center}
\epsfig{file=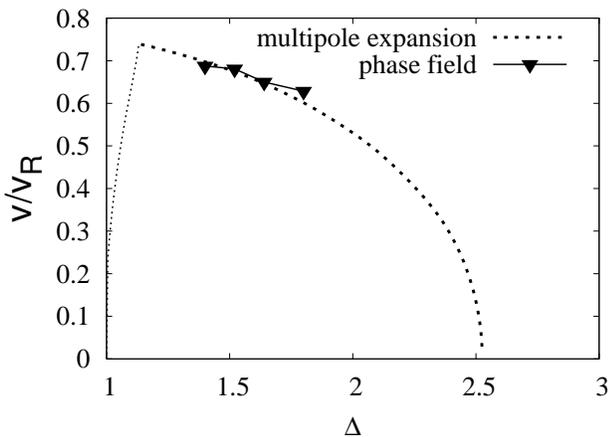, angle=-90, width=8cm}
\caption{Comparison of the steady-state crack velocity obtained from the multipole expansion technique \cite{Pilipenko06} and the extrapolated value from phase field simulations.}
\label{VelocityComparison}
\end{center}
\end{figure}

This tedious procedure was performed for several driving forces, and in Fig.~\ref{VelocityComparison} the comparison to the multipole expansion method \cite{Pilipenko06} is shown.
The agreement of the results which are obtained from completely different methods is very convincing.
The small deviation for $\Delta=1.8$ is due to the fact that this value is already close to the threshold of the tip-splitting instability which cannot be captured by the multipole expansion method.
In particular, we find evidence for the prediction that the steady state velocity decays weakly with increasing driving force, which might be an artefact of the model.
%Notice that the interval of steady state solution and velocities are rather small, as for $\Delta\geq 1.9$ tip splitting occurs.

\begin{figure}
\begin{center}
\epsfig{file=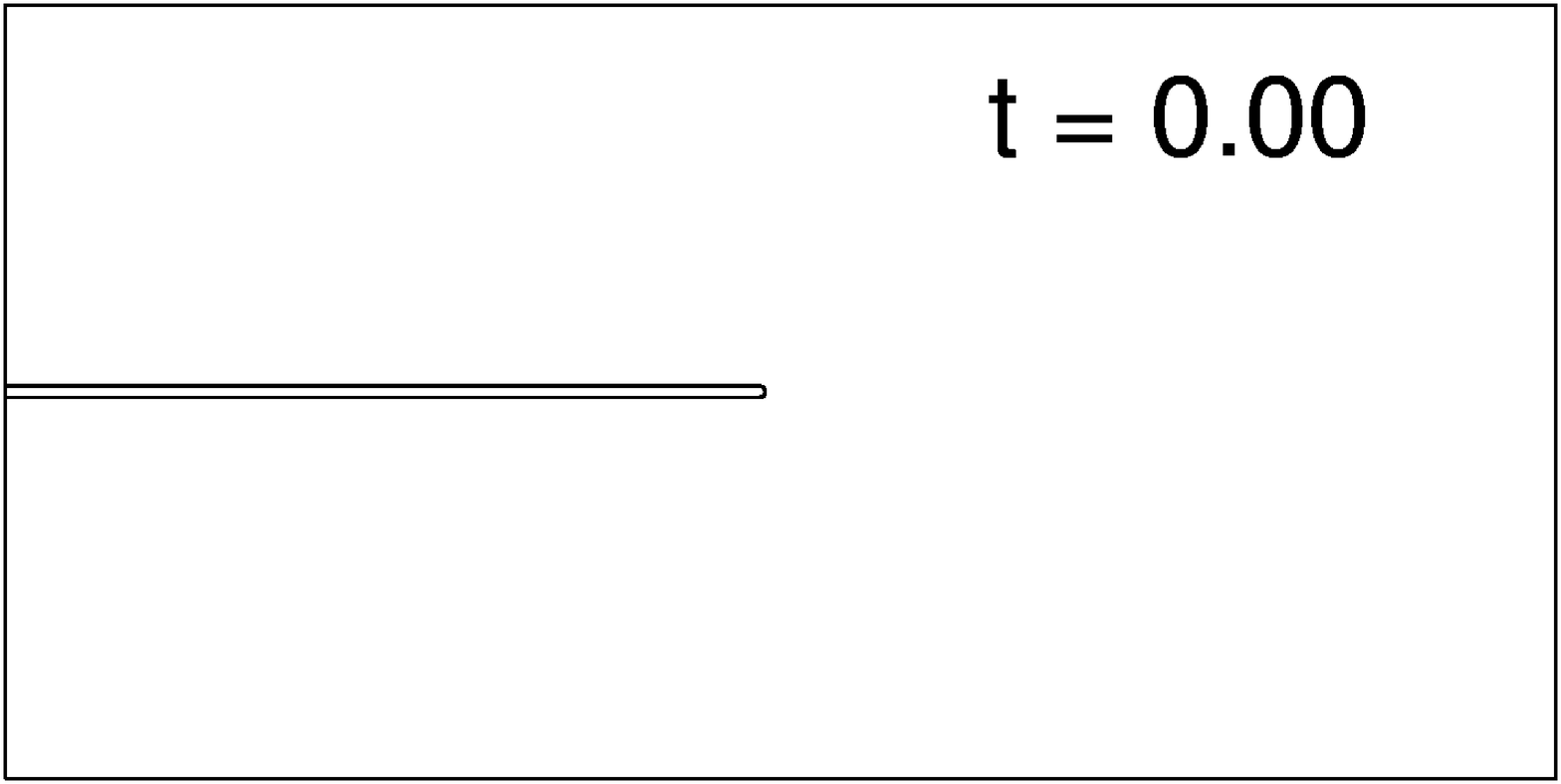, width=4cm}
\epsfig{file=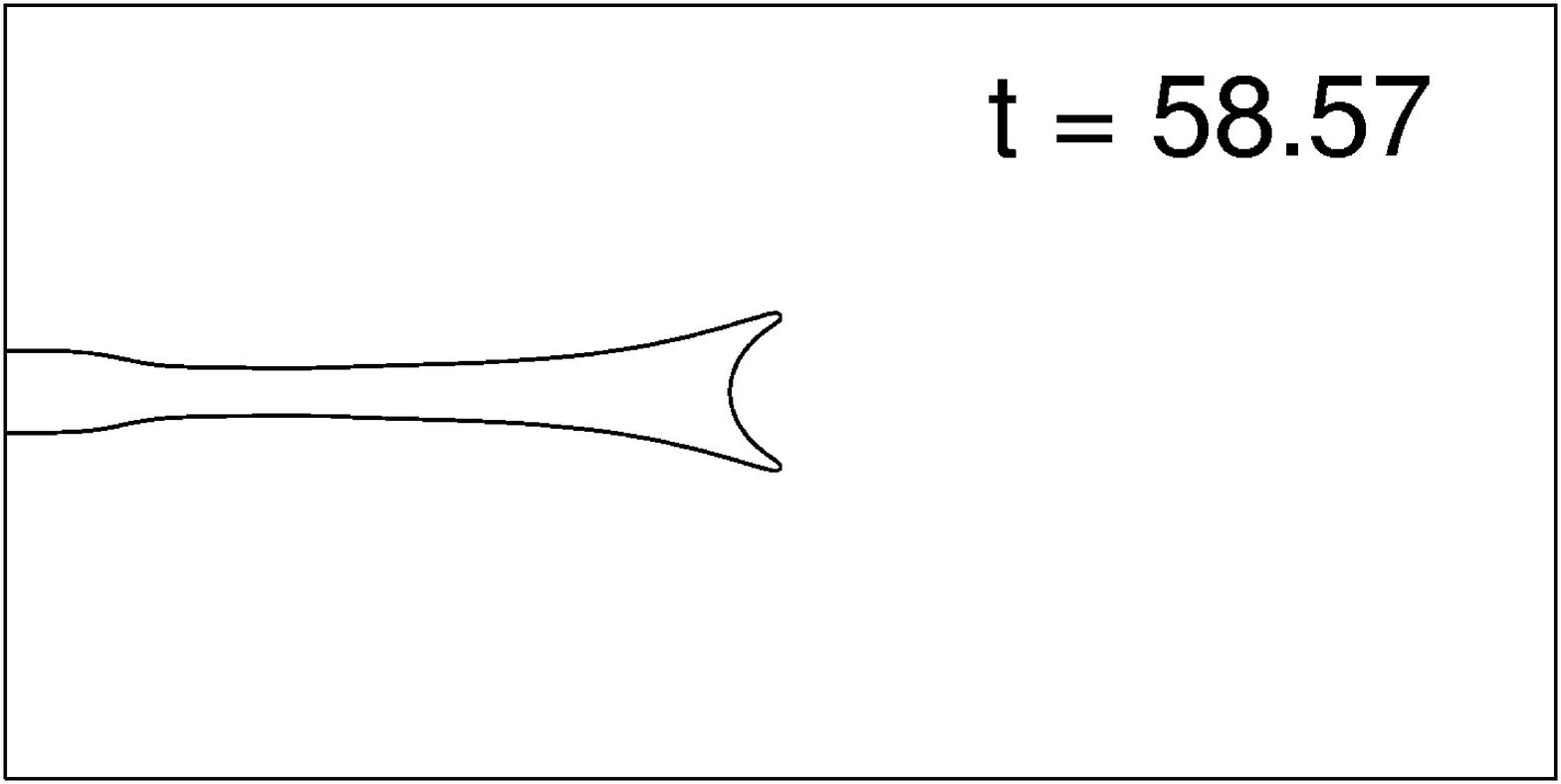, width=4cm} \\[0.3ex]
\epsfig{file=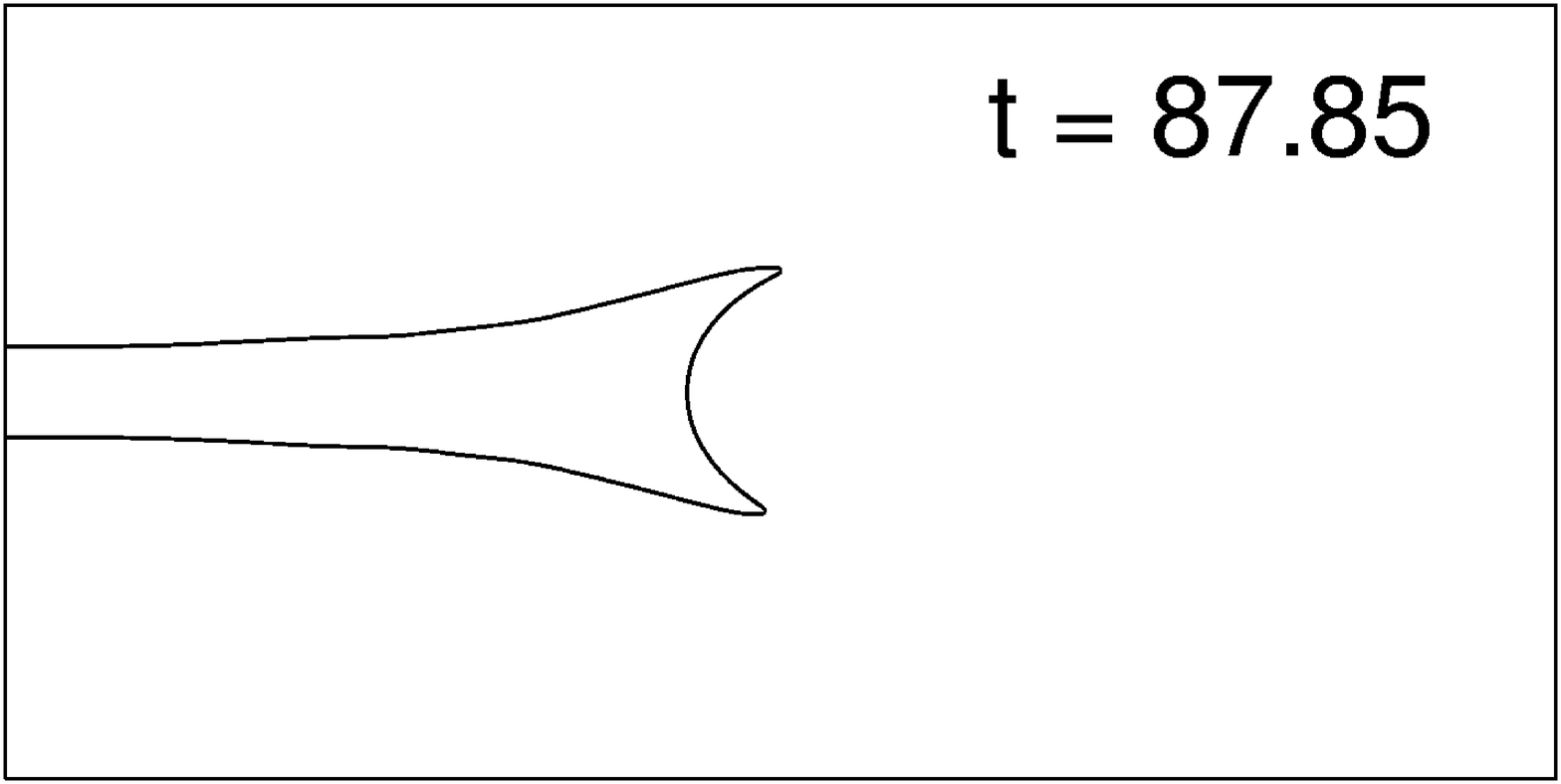, width=4cm}
\epsfig{file=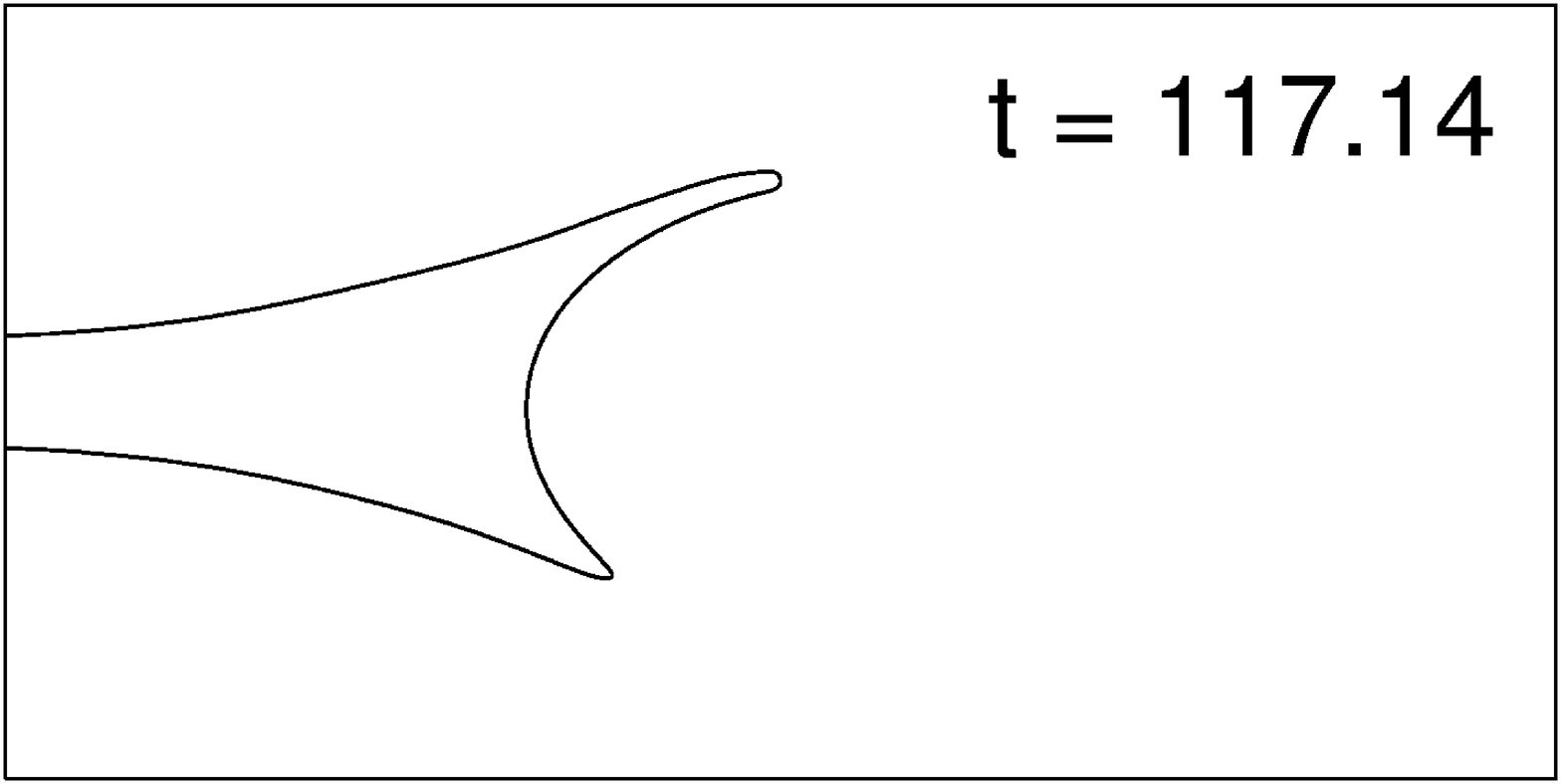, width=4cm}\\[0.3ex]
\epsfig{file=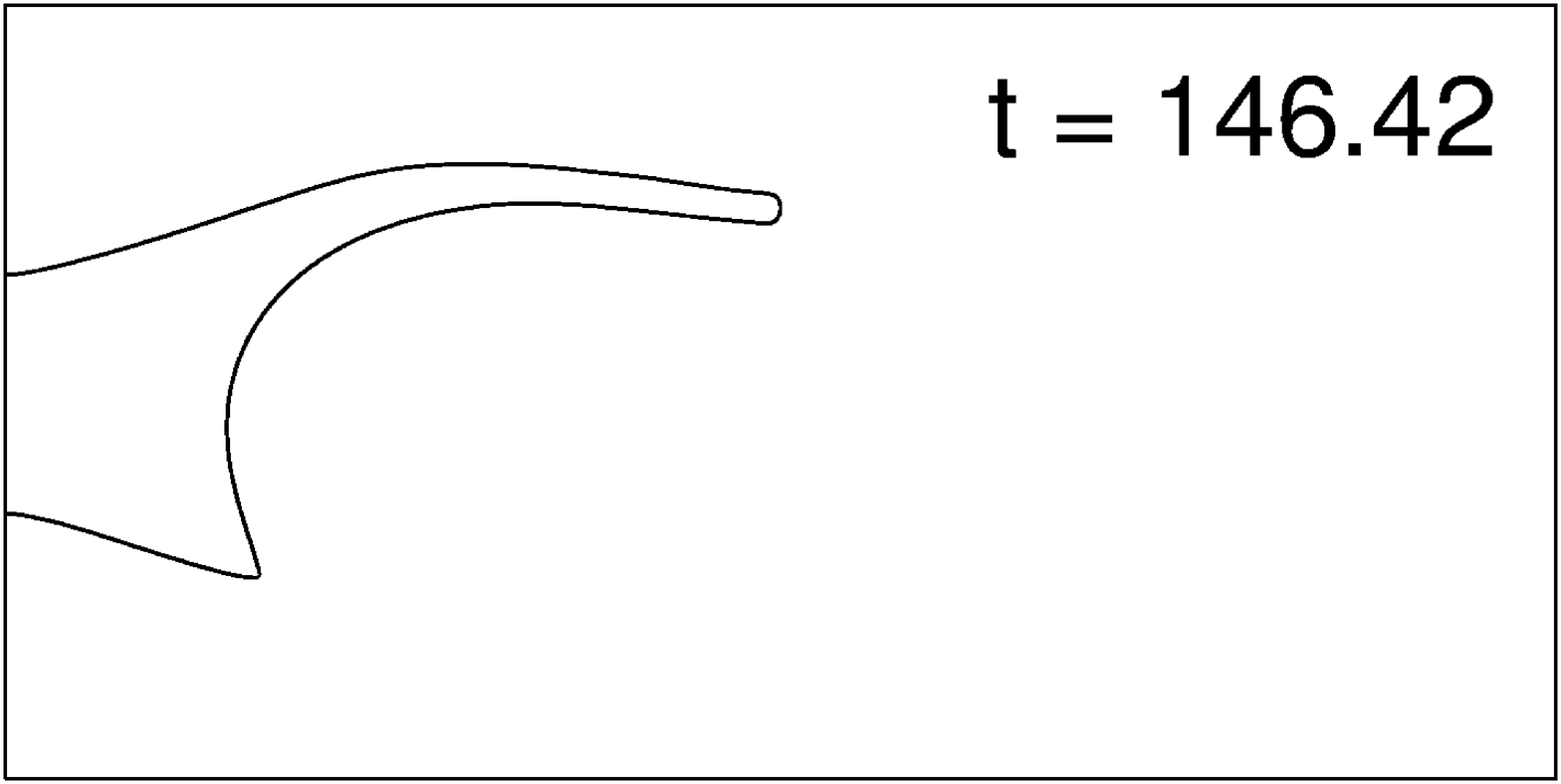, width=4cm}
\epsfig{file=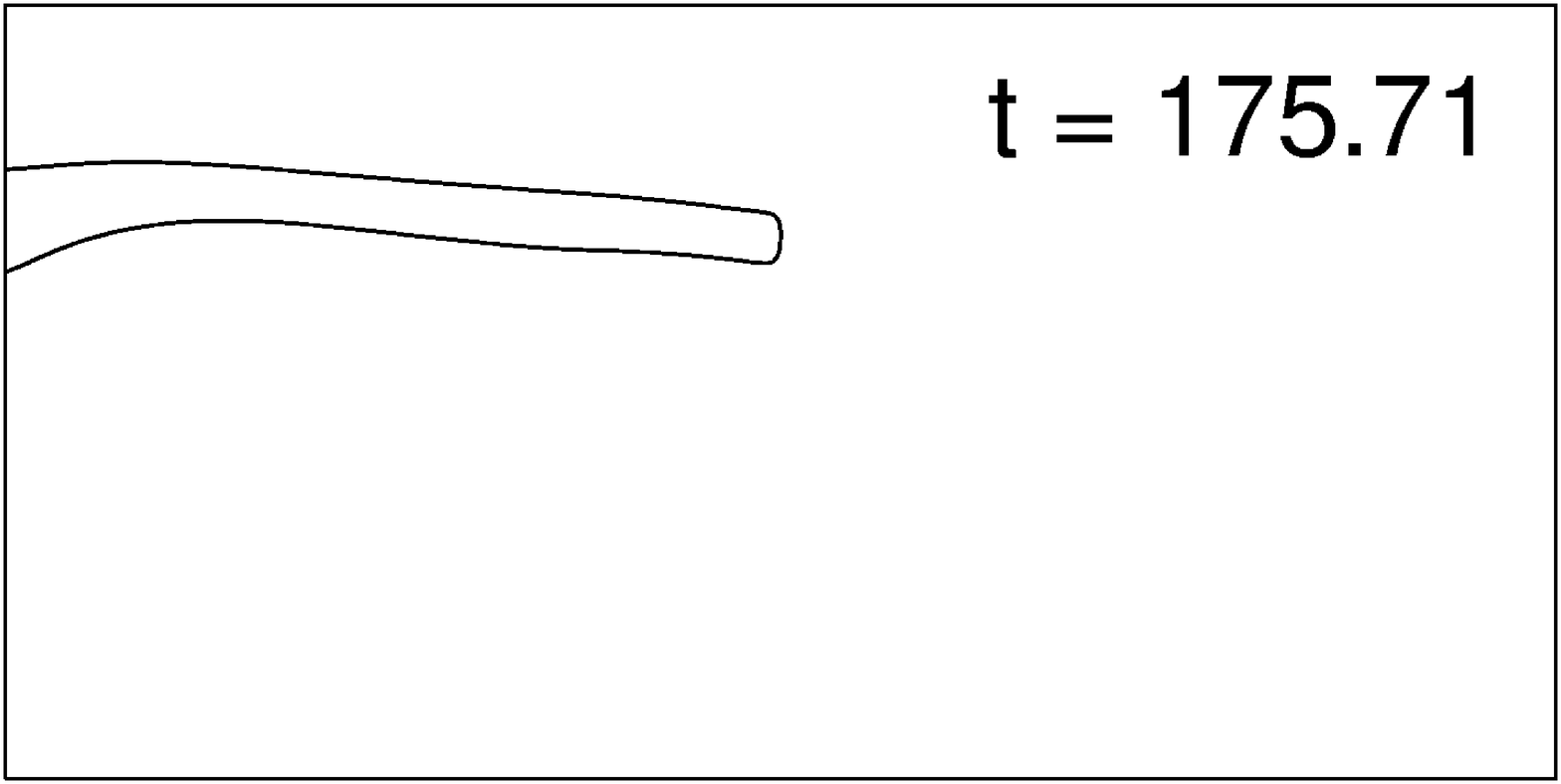, width=4cm}
\caption{Irregular tip splitting scenario for $\Delta=1.9$. We used $Lv_R/D=44.2$ and $D/v_R \xi=9.3$;
the aspect ratio of the system is 2:1.
Time is given in units $D/v_R^2$.
The thickness of the interface is the phase field interface width, indicating a good separation of the scales.}
\label{splitfig}
\end{center}
\end{figure}

For higher driving forces, we observe tip-splitting in the phase-field simulations (see Fig.~\ref{splitfig}).
The onset of this irregular behavior depends sensitively on the system size, because in relatively small systems, the branches of the crack cannot separate since they are repelled by the boundaries.
Therefore, the steady state growth is always stabilized by finite size effects.
On the other hand, initial conditions can trigger an instability, and then a long transient is required to get back to steady state solutions.
Despite these restrictions, we are still able to make the prediction that the threshold of splitting obeys $\Delta_{split}\lesssim 1.9$ in the phase field model.
It is in agreement with the conjecture that branching occurs as soon as the steady state tip curvature becomes negative, leading to the prediction $\Delta_{split}\approx 1.8$ \cite{Pilipenko06}.

The numerical determination of a characteristic crack width scale in the sharp interface limit is more difficult, and therefore we refrain from performing a double extrapolation procedure.
The explanation is that if the soft phase inside the crack still possesses small nonvanishing elastic constants, the equilibrium situation far behind the crack tip corresponds to a full opening of the crack, instead of the opening being of the order $D/v_R$:
As it is shown in Appendix \ref{planarAppendix}, the elastic energy is minimized if the hard phase completely disappears.
Small remaining elastic constants can be due to an insufficient separation of the scales $D/v_R$ and $\xi$, since according to Eqs.~(\ref{muinterpolated}) and (\ref{lambdainterpolated}), the elastic constants decay only exponentially inside the crack, even if this soft phase has nominally vanishing elastic coefficients.
Therefore, the crack opening is a weakly growing function of the distance from the crack tip, and this slope becomes smaller with better scale separation, see Fig.~\ref{CrackOpenings}.
We point out that this opening is solely due to the phase transition process, and the shapes are drawn without elastic displacements which should be added to obtain the real shape under load.
For example, the vertical displacement obeys the usual scaling $u_y\sim \sqrt{|x|}$ for large distances $|x|$ from the tip.

\begin{figure}
\begin{center}
\epsfig{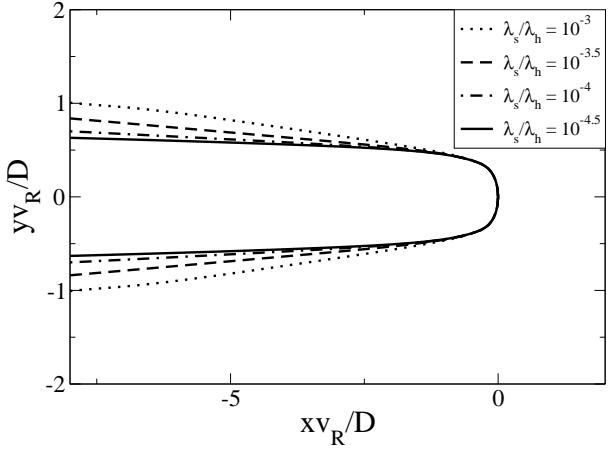}
\caption{Solid-solid transformation in a strip. A very soft phase grows at the expense of a harder phase. Parameters are $Lv_R/D=11.03$ ($v_R$ is the Rayleigh speed of the hard phase), $D/\xi v_R=9.27$, the aspect ratio is $2:1$. The Poisson ratio $\nu=1/3$ is equal in both phases, and the driving force is $\Delta=1.4$.}
\label{SoftPhasesFig}
\end{center}
\end{figure}

The same effect can be seen if we investigate solid-solid transformations towards a soft phase with small elastic constants.
The Poisson ratios in both the surrounding solid and the new inner phase are chosen equally, $\nu=1/3$, but the bulk moduli differ by many orders of magnitude.
The softer the inner phase becomes, the less the opening of the ``crack'' grows with increasing distance from the tip, see Fig.~\ref{SoftPhasesFig}.
Only very far away, the new phase fills the whole channel.

%%%%%%%%%%%%%%%%%%%%%%%%%%%%%%%%%%%%%%%%%%%%%%%%%%%%%%%%%%%%%%%%%%%%%
\section{Multiphase modeling}
\label{multiphase}

A simple approach to derive multiphase equations to describe systems consisting of more than two phases starts again from variational principles.
The volume fraction of each phase is described by a field variable $\phi_k$, $k=1,\ldots,N$ with $N$ being the number of phases, $\Phi=(\phi_1,\ldots,\phi_N)$.
In the sharp interface limit, one phase field variable has the value one inside the bulk phases, and the others are zero.
Their temporal evolution is given by
\begin{equation}
\frac{\partial\phi_k}{\partial t} = - \frac{2\tilde{D}}{3\xi} \left( \frac{\delta F}{\delta \phi_k} -\Lambda \right),
\end{equation}
where we introduce a Lagrange multiplier to maintain the phase conservation, $\sum_{k=1}^N \phi_k=1$.
We redefine the diffusion coefficient $D/\gamma \to \tilde{D}$, which is more appropriate to generalize to arbitrary interfacial energy coefficients $\gamma_{ik}$ between phases $i$ and $k$.
%Notice that an additional factor of two has appeared in the equation of motion above, which is related to the Lagrange multiplier.
The additional factor $2$ in the equation of motion above is chosen to recover the previous phase field model in the case $N=2$.
The expression for the Lagrange multiplier is given by
\begin{equation}
\Lambda = \frac{1}{N} \sum_{i=1}^N \frac{\delta F}{\delta \phi_k}.
\end{equation}
The free energy $F=F_{el}+F_s+F_{dw}$ has the following contributions:
\begin{equation}
F_{el}= \int \left( \mu(\Phi) \epsilon_{ij}^2 + \lambda(\Phi)(\epsilon_{ii})^2/2\right) dV,
\end{equation}
\begin{equation}
F_s = \frac{3\xi}{4}\sum_{i,j=1}^N \gamma_{ij} \int (\phi_i\nabla\phi_j-\phi_j\nabla\phi_i)^2 dV,
\end{equation}
\begin{equation} \label{multiwell}
F_{dw} = \frac{3}{\xi} \sum_{{i,j=1 \atop i\neq j}}^N \gamma_{ij}\int \phi_i^2\phi_j^2 dV.
\end{equation}
Here, the interpolated elastic constants are
\begin{equation}
\mu(\Phi) = \sum_{i=1}^N h(\phi_i) \mu^{(i)}, \qquad \lambda(\Phi) = \sum_{i=1}^N h(\phi_i) \lambda^{(i)},
\end{equation}
with $\mu^{(i)}$ and $\lambda^{(i)}$ being the elastic constants of the individual bulk phases.
Also, we have mutual interfacial energy coefficients $\gamma_{ij}=\gamma_{ji}$ for phase boundaries between $i$ and $j$.
Notice that third-phase contributions appear inside two-phase boundaries which lead to a renormalization of the bare interfacial energies.
This can be avoided by the addition of higher order terms to the multiwell-potential (\ref{multiwell}).
See Refs.~\cite{Nestler,Garcke} for a discussion of this issue.

Variation gives explicitly:
\begin{eqnarray*}
\frac{\delta F_{el}}{\delta \phi_k} &=& h'(\phi_k) \left( \mu_k \epsilon_{ij}^2 + \lambda_k(\epsilon_{ii})^2/2 \right), \\
\frac{\delta F_s}{\delta\phi_k} &=& 3\xi \sum_{i=1}^N \gamma_{ik} \Big[ 2(\phi_k\nabla\phi_i - \phi_i\nabla\phi_k)\cdot\nabla\phi_i \\
&& - \phi_i(\phi_i\nabla^2\phi_k-\phi_k\nabla^2\phi_i) \Big], \\
\frac{\delta F_{dw}}{\delta\phi_k} &=& \frac{12}{\xi} \sum_{{i=1 \atop i\neq k}}^N \gamma_{ik}\phi_i^2\phi_k.
\end{eqnarray*}
Similarly, the elastic equation of motion is just the generalization of Eq.~(\ref{phase:eq2})
\begin{equation}
\rho \ddot{u}_i = \frac{\partial \sigma_{ik}(\Phi)}{\partial x_k}
\end{equation}
with
\begin{equation}
\sigma_{ik}(\Phi) = 2\mu(\Phi)\epsilon_{ik} + \lambda(\Phi)\epsilon_{ll}\delta_{ik}.
\end{equation}
This description is a direct generalization of Ref.~\cite{Nestler}, and therefore it is known that it leads to appropriate sharp interface equations similar to Eqs.~(\ref{coherency})-(\ref{normalvel}) and contact angles as predicted by Young's law \cite{Garcke}.

%%%%%%%%%%%%%%%%%%%%%%%%%%%%%%%%%%%%%%%%%%%%%%%%%%%%%%%%%%%%%%%%%%%%%
%%%%%%%%%%%%%%%%%%%%%%%%%%%%%%%%%%%%%%%%%%%%%%%%%%%%%%%%%%%%%%%%%%%%%
\appendix

%%%%%%%%%%%%%%%%%%%%%%%%%%%%%%%%%%%%%%%%%%%%%%%%%%%%%%%%%%%%%%%%%%%%%
\section{Derivation of the sharp interface equations}
\label{SharpInterfaceAppendix}

Here we explain in detail how the equations of motion, the appropriate boundary conditions and the chemical potential, which is responsible for interface motion, are derived in a unique way from variational principles.
We assume that the two phases are coherent, i.e.~the displacement field is continuous across the interface, and the mass densities are equal in both phases.
Since we do not consider lattice strains or surface tension here, all elastic stresses arise from external forces.
For simplicity, we assume a two-dimensional plane-strain situation.
Contributions due to surface energy do not couple to the elastic fields.
Ultimately, they give only an additive curvature-dependent term to the chemical potential, as incorporated in Eq.~(\ref{normalvel}).
We note that this expression was already obtained in \cite{Privorotskii}, but here we take also inertial effects into account, since the velocity of cracks is typically of the order of the sound speeds.

The kinetic energy density is in both phases
\begin{equation}
T = \frac{1}{2} \rho \dot{u}_i^2,
\end{equation}
and the potential energy density reads
\begin{equation}
U^{(\alpha)} = \frac{1}{2}\sigma_{ik}^{(\alpha)} \epsilon_{ik}^{(\alpha)}.
\end{equation}
Notice that certain components of the stress and strain tensors are in general discontinuous at the interface, as will be elaborated below.

We assume the total volume $V$ of the entire system to be constant in time and to be decomposed into two subvolumes $V^{(1)}(t)$ and $V^{(2)}(t)$ of different solids.
Upper indices discriminate between the phases (see Fig.~\ref{fig1}).
The common interface $A(t) := \partial V^{(1)}(t) \cap \partial V^{(2)}(t)$ with normal $n$ and tangential $\tau$ is moving in time due to phase transitions, and consequently, the phase volumes are time-dependent as well.
However, we do not yet specify a concrete dynamical process here.

The Lagrangian is defined as
\begin{equation}
L(t) = \int\limits_V T dV - \int\limits_{V_1(t)} U^{(1)} dV - \int\limits_{V_2(t)} U^{(2)} dV,
\end{equation}
and the action is
\begin{equation} \label{action}
S  = \int\limits_{t_0}^{t_1} L(t) dt,
\end{equation}
with arbitrary beginning and end times $t_0$ and $t_1$.

We obtain the usual elastic equations by varying the action (\ref{action}) with respect to the displacement field for fixed interface positions.
%In contrast, in the next section, we will obtain the chemical potential and the equation of motion for the phase transition by variation of the interface profile for fixed elastic fields.
Thus we get
\begin{eqnarray*}
\delta S &=& \int\limits_{t_0}^{t_1} dt \Bigg[ \int\limits_V \rho \dot{u}_i\delta\dot{u}_i dV - \int\limits_{V_1(t)} \sigma_{ik}^{(1)}\delta \epsilon_{ik}^{(1)} dV \\
&& -\int\limits_{V_2(t)} \sigma_{ik}^{(2)}\delta \epsilon_{ik}^{(2)} dV \Bigg] \\
&=& \int\limits_{t_0}^{t_1} dt \Bigg[ \int\limits_V \rho \dot{u}_i\delta\dot{u}_i dV + \int\limits_{V_1(t)} \frac{\partial\sigma_{ik}^{(1)}}{\partial x_k} \delta u_i dV \\
&&- \int\limits_{A(t)} \sigma_{in^{(1)}}^{(1)}\delta u_i d\tau + \int\limits_{V_2(t)} \frac{\partial\sigma_{ik}^{(2)}}{\partial x_k} \delta u_i dV \\
&&- \int\limits_{A(t)} \sigma_{in^{(2)}}^{(2)}\delta u_i d\tau \Bigg].
\end{eqnarray*}
The first integral is integrated by parts, assuming as usual that the variations $\delta u_i$ vanish for $t_0$ and $t_1$.
Since also the normal vectors of both phases are antiparallel, ${\bf n} := {\bf n}^{(1)}=- {\bf n}^{(2)}$, thus $\sigma_{in^{(2)}}=-\sigma_{in}$, we get
\begin{eqnarray*}
\delta S &=& \int\limits_{t_0}^{t_1} dt \Bigg[ \int\limits_{V_1(t)} \left( \frac{\partial\sigma_{ik}^{(1)}}{\partial x_k} - \rho \ddot{u}_i \right) \delta u_i dV \\
&& + \int\limits_{V_2(t)} \left( \frac{\partial\sigma_{ik}^{(2)}}{\partial x_k} -\rho \ddot{u}_i \right) \delta u_i dV \\
&& - \int\limits_{A(t)} (\sigma_{in}^{(1)}-\sigma_{in}^{(2)}) \delta u_i d\tau \Bigg].
\end{eqnarray*}
Demanding vanishing variation $\delta S$ gives in the bulk the usual equations of motion
\begin{equation} \label{dynelastAppendix}
\frac{\partial \sigma_{ik}^{(\alpha)}}{\partial x_k} = \rho \ddot{u}_i,
\end{equation}
and on the interface we obtain the continuity of normal and shear stresses
\begin{equation} \label{boundaryAppendix}
\sigma_{in}^{(1)} = \sigma_{in}^{(2)}.
\end{equation}

The next step is to calculate the change of the total energy when the interface moves in the course of time.
This is done in three steps:
First, we calculate the change of energy due to the time evolution of the elastic fields for fixed interface position.
Second, we calculate the change of elastic energy due to the motion of the interface for fixed elastic fields in the bulk phases.
After this second step, the coherency condition at the interface is violated.
In the last step, we therefore have to do additional work to adjust the displacements appropriately.

The first contribution is
\begin{eqnarray*}
\frac{d W_1}{dt} &=& \int\limits_{V^{(1)}(t)} \frac{\partial}{\partial t} (T+U^{(1)}) dV \\
&+&  \int\limits_{V^{(2)}(t)} \frac{\partial}{\partial t} (T+U^{(2)}) dV \\
&=& \int\limits_V \rho \dot{u}_i\ddot{u}_i dV + \int\limits_{V^{(1)}(t)} \sigma_{ik}^{(1)} \dot{\epsilon}_{ik}^{(1)} dV \\
&+&  \int\limits_{V^{(2)}(t)} \sigma_{ik}^{(2)} \dot{\epsilon}_{ik}^{(2)} dV.
\end{eqnarray*}
We note that the kinetic energy density is continuous across the interface.
Furthermore, by the equations of motion (\ref{dynelastAppendix})
\begin{eqnarray*}
\frac{d W_1}{dt} &=& \int\limits_{V^{(1)}(t)} \dot{u}_i \frac{\partial\sigma_{ik}^{(1)}}{\partial x_k} dV + \int\limits_{V^{(2)}(t)} \dot{u}_i \frac{\partial\sigma_{ik}^{(2)}}{\partial x_k} dV \\
&+& \int\limits_{V^{(1)}(t)} \frac{\partial}{\partial x_k} \left( \sigma_{ik}^{(1)} \dot{u}_i \right) dV - \int\limits_{V^{(1)}(t)} \frac{\partial \sigma_{ik}^{(1)}}{\partial x_k} \dot{u}_i dV \\
&+& \int\limits_{V^{(2)}(t)} \frac{\partial}{\partial x_k} \left( \sigma_{ik}^{(2)} \dot{u}_i \right) dV - \int\limits_{V^{(2)}(t)} \frac{\partial \sigma_{ik}^{(2)}}{\partial x_k} \dot{u}_i dV \\
&=& \int\limits_{A(t)} \sigma_{in^{(1)}}^{(1)} \dot{u}_i d\tau + \int\limits_{A(t)} \sigma_{in^{(2)}}^{(2)} \dot{u}_i d\tau = 0,
\end{eqnarray*}
where we assumed for simplicity that $\dot{u}_i=0$ on all boundaries apart from $A(t)$, i.~e.~no external work is exerted to the solids.
In the last step, we used the boundary conditions (\ref{boundaryAppendix}), $\sigma^{(1)}_{in} = \sigma^{(2)}_{in} = -\sigma^{(2)}_{in^{(2)}}$;
also, by definition, the displacement rate $\dot{u}_i$ is continuous across the interface.
The above result is quite clear since the elastodynamic time evolution is purely conservative.

The second contribution arises due to the motion of the interface for fixed elastic fields.
We extend the elastic state of the growing phase analytically into the newly acquired region.
This assures that the bulk equations remain fulfilled in both phases even after the forward motion of the interface.
Thus this contribution to the energy change rate reads
\begin{equation}
\frac{d W_2}{dt} = \int\limits_{A(t)} v_n \left( U^{(1)} - U^{(2)} \right) d\tau.
\end{equation}
The interface normal velocity is positive if the phase 1 locally extends.
Here, we immediately used the continuity of the kinetic energy density, which therefore cancels.

After the phase transformation in this second step, the displacements are no longer continuous at the interface.
Thus extra work has to be invested to remove this misfit.
%We introduce a local coordinate system $n$ and $\tau$ for the perpendicular normal and tangential directions (see Fig.~\ref{fig1}).
%Then the strain tensor becomes
In the local coordinate system $n$ and $\tau$ (see Fig.~\ref{fig1}) the strain tensor becomes
\begin{eqnarray}
\epsilon_{nn} &=& \partial_n u_n, \\
\epsilon_{\tau\tau} &=& \partial_\tau u_\tau + \kappa u_n, \\
\epsilon_{n\tau} = \epsilon_{\tau n} &=& \frac{1}{2} \left( \partial_\tau u_n + \partial_n u_\tau - \kappa u_\tau \right).
\end{eqnarray}
Here, $\kappa$ is the interface curvature, which is positive if the phase 1 is convex.

At this point, a few comments concerning the continuity of various fields across the coherent interface are in order.
Since the displacement field has to be continuous across the interface, also its tangential derivatives are continuous, but the normal derivatives are not.
Consequently, the following quantities are continuous: $\partial_\tau u_\tau, \partial_\tau u_n, \kappa u_n, \kappa u_\tau, \epsilon_{\tau\tau}$.
On the other hand, $\partial_n u_n, \partial_n u_\tau, \epsilon_{nn}, \epsilon_{n\tau}$ are discontinuous across the interface.

In the second step of energy calculation, we extended smoothly the fields into the receding domain.
The interface at this new time $t+\Delta t$ is now located at a different position.
This leads to discontinuities of the displacements, e.~g.~for the normal component at the new position of the interface
\[
\Delta u_n = \left( [\partial_n u_n]^{(1)} -  [\partial_n u_n]^{(2)} \right) v_n \Delta t= (\epsilon_{nn}^{(1)} - \epsilon_{nn}^{(2)}) v_n \Delta t,
\]
where $[\ldots]^{(\alpha)}$ denotes the evaluation of a probably discontinuous expression at the previous interface position, taken for the phase $\alpha$.
Similarly, for the tangential component
\begin{eqnarray*}
\Delta u_\tau &=& \big( 2\epsilon_{n\tau}^{(1)} - [\partial_\tau u_n]^{(1)} + [\kappa u_\tau]^{(1)} - 2\epsilon_{n\tau}^{(2)} + [\partial_\tau u_n]^{(2)} \\
&&- [\kappa u_\tau]^{(2)} \big) v_n \Delta t \\
&=& 2(\epsilon_{n\tau}^{(1)} - \epsilon_{n\tau}^{(2)}) v_n \Delta t.
\end{eqnarray*}
To zeroth order in $\Delta t$, the stresses at the new interface position are equal on both sides and identical to the stresses at the previous interface position.
To reconnect the displacements, we have to apply the coherency work rate
\[
\frac{d W_3}{dt} = \int\limits_{A(t)} v_n \left[ -(\epsilon_{nn}^{(1)} - \epsilon_{nn}^{(2)}) \sigma_{nn} - 2 (\epsilon_{n\tau}^{(1)} - \epsilon_{n\tau}^{(2)}) \sigma_{n\tau} \right] d\tau.
\]
Altogether, the change of the energy is given by
\begin{eqnarray*}
\frac{dW}{dt} &=& \frac{d(W_1+W_2+W_3)}{dt} \\
&=& \int\limits_{A(t)} v_n \Bigg[ \left( \frac{1}{2} \sigma_{\tau\tau}^{(1)}\epsilon_{\tau\tau}^{(1)} - \frac{1}{2} \sigma_{nn}^{(1)}\epsilon_{nn}^{(1)} - \sigma_{n\tau}^{(1)}\epsilon_{n\tau}^{(1)} \right) \\
&& - \left( \frac{1}{2} \sigma_{\tau\tau}^{(2)}\epsilon_{\tau\tau}^{(2)} - \frac{1}{2} \sigma_{nn}^{(2)}\epsilon_{nn}^{(2)} - \sigma_{n\tau}^{(2)}\epsilon_{n\tau}^{(2)} \right) \Bigg] d\tau.
\end{eqnarray*}
We  can therefore define an appropriate chemical potential for each phase at the coherent interface
\begin{equation}
\mu^{(\alpha)}_{el} = \Omega \left( \frac{1}{2} \sigma_{\tau\tau}^{(\alpha)}\epsilon_{\tau\tau}^{(\alpha)} - \frac{1}{2} \sigma_{nn}^{(\alpha)}\epsilon_{nn}^{(\alpha)} - \sigma_{n\tau}^{(\alpha)}\epsilon_{n\tau}^{(\alpha)} \right).
\end{equation}
Notice that, in contrast to a free surface, the normal and shear contributions appear with negative sign.
Then, the energy dissipation rate can be written as
\begin{equation}
\frac{dW}{dt} = \frac{1}{\Omega}\int\limits_{A(t)} (\mu^{(1)}_{el} - \mu^{(2)}_{el}) v_n d\tau.
\end{equation}
Due to the coherency condition and the requirement of equal mass density, the kinetic energy density does not appear.

%This derivation of the chemical potential is based rather on physical arguments;
%a more formal derivation will be given elsewhere \cite{Fleck}.

%%%%%%%%%%%%%%%%%%%%%%%%%%%%%%%%%%%%%%%%%%%%%%%%%%%%%%%%%%%%%%%%%%%%%
\section{Motion of a planar phase boundary}
\label{planarAppendix}

The elastic contribution to the chemical potential (\ref{chempot}) at the interface between two solid phases has the remarkable property that the normal and shear contributions are negative definite.
This means that growth of the phase with higher elastic energy density at the expense of the phase with lower energy density can still reduce the total energy.
In order to illuminate this point, we consider a simple example.

\begin{figure}
\begin{center}
\epsfig{file=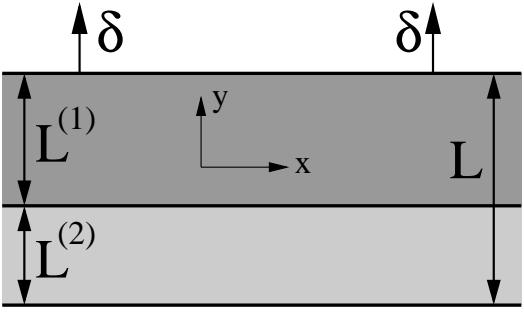, width=7cm}
\caption{Motion of a planar interface in a strip geometry. Vertically, a constant displacement $\delta$ is applied to the grips of the strip which consists of two solid phases with different elastic constants.}
\label{fig2}
\end{center}
\end{figure}

Here, two strips of different solid materials are coherently connected (see Fig.~\ref{fig2}), and the interface can move due to a phase transition.
We assume for simplicity that the process is slow and inertial effects can be neglected.
We apply a fixed displacement $\delta$ at the upper end of this layer structure and set the displacement at the lower grip to zero.
The total strip width $L$ is distributed among the two layers, $L=L^{(1)} + L^{(2)} = const$.
We have a homogeneous strain situation in each of the phases ($\alpha=1,2$), i.e. $u_x^{(\alpha)}=0$, $u_y^{(\alpha)}=u_y^{(\alpha)}(y)$ with a pure linear dependence of $u_y^{(\alpha)}$ on $y$.
Strains are $\epsilon_{xx}^{(\alpha)}=\epsilon_{xy}^{(\alpha)}=0$, $\epsilon_{yy}=\partial_y u_y^{(\alpha)}=const$.
The elongation of each phase is $\delta^{(\alpha)}=L^{(\alpha)} \epsilon_{yy}^{(\alpha)}$.
In sum, they are equal to the prescribed total opening $\delta=\delta^{(1)}+\delta^{(2)}=const$.
Stresses and strain are connected through Hooke's law for isotropic materials, thus $\sigma_{xx}^{(\alpha)} = \lambda^{(\alpha)} \epsilon_{yy}^{(\alpha)}$, $\sigma_{xy}^{(\alpha)}=0$, $\sigma_{yy}^{(\alpha)} = (2\mu^{(\alpha)} + \lambda^{(\alpha)}) \epsilon_{yy}^{(\alpha)}$.
At the interface, the equality of normal stresses, $\sigma_{yy}^{(1)} = \sigma_{yy}^{(2)}$, leads to
\begin{equation}
\epsilon_{yy}^{(2)} = \frac{2\mu^{(1)} + \lambda^{(1)}}{2\mu^{(2)} + \lambda^{(2)}} \epsilon_{yy}^{(1)}.
\end{equation}
The strain can be computed in terms of the given total opening,
\begin{equation}
\epsilon_{yy}^{(1)} = \frac{\delta}{L^{(1)} + (L-L^{(1)}) \frac{2\mu^{(1)} + \lambda^{(1)}}{2\mu^{(2)} + \lambda^{(2)}}}.
\end{equation}
Hence, we can calculate the total elastic energy in the strip (per unit length) as function of $L^{(1)}$:
\begin{eqnarray}
{\cal U}(L^{(1)}) &=& \frac{1}{2} L^{(1)} \sigma_{ik}^{(1)} \epsilon_{ik}^{(1)} + \frac{1}{2} L^{(2)} \sigma_{ik}^{(2)} \epsilon_{ik}^{(2)} \nonumber \\
&=& \frac{1}{2} \frac{\delta^2 (2\mu^{(1)}+\lambda^{(1)})}{L^{(1)} + (L-L^{(1)}) \frac{2\mu^{(1)}+\lambda^{(1)}}{2\mu^{(2)}+\lambda^{(2)}}}. \label{stripenergy}
\end{eqnarray}
This function is monotonic in $L^{(1)}$.
Assuming that phase 1 is harder than phase 2, $2\mu^{(1)}+\lambda^{(1)} > 2\mu^{(2)}+\lambda^{(2)}$, the energy is minimized if $L^{(1)}=0$, i.e.~if the hard phase disappears.
Notice that on the other hand, the elastic energy density is higher in the softer phase, $\sigma_{ik}^{(1)} \epsilon_{ik}^{(1)}/2 < \sigma_{ik}^{(2)} \epsilon_{ik}^{(2)}/2$.
We get from Eq.~(\ref{stripenergy})
\[
\frac{d{\cal U}^{(1)}}{dL^{(1)}} = -\frac{1}{2} \sigma_{yy}^{(1)}\epsilon_{yy}^{(1)} + \frac{1}{2} \sigma_{yy}^{(2)}\epsilon_{yy}^{(2)} = \frac{1}{\Omega}\left( \mu^{(1)}_{el} - \mu^{(2)}_{el} \right).
\]
Here we clearly see that the {\em negative} elastic energy density for the normal direction enters into the energy change rate and thus into the chemical potential;
in more general cases one can easily verify that this is also true for shear contributions.

%%%%%%%%%%%%%%%%%%%%%%%%%%%%%%%%%%%%%%%%%%%%%%%%%%%%%%%%%%%%%%%%%%%%%
\section{Phase field modeling of coherent solids}
\label{CoherentPhasefieldAppendix}

The aim of this section is to show that the phase field model presented in Section \ref{PhasefieldSection} leads to the correct sharp interface limit even for solid-solid transformations, where the chemical potential differs from the usual expression \cite{Nozieres93}.
Since the treatment of the surface energy contribution is well-known and enters additively into the chemical potential, we focus on the elastic fields here.

For smooth interfaces, all stresses remain finite even in the sharp interface limit.
We note that this statement holds even for fracture processes, where usually stresses can diverge at sharp corners and tips in the framework of the linear theory of elasticity;
nevertheless, in the current description, the tip radius $r_0$ is always finite and therefore stresses are limited to values $\sigma\sim K r_0^{-1/2}$, where $K$ is a stress intensity factor.
Consequently, the displacement field must be continuous in the sharp interface limit, because a finite mismatch $\delta u$ would lead to diverging stresses and strains $\epsilon\sim \sigma/E\sim \delta u/\xi$, and thus a divergent elastic energy.
Then, obviously, the equation of motion (\ref{phase:eq2}) leads to the usual elastic bulk equations and also to the continuity of normal and shear stresses at the interface in the limit $\xi\to 0$.

Next, we calculate the total elastic energy change due to the interface motion.
We determine the energy contributions and changes in a line perpendicular through the interface, assuming the interface curvature to be small, $\kappa\xi \ll 1$.
To do so, we introduce a local coordinate $n$ normal to the interface (pointing from phase 1 into phase 2).
The total elastic energy per unit width is defined as
\[
{\cal W} = \int\limits_{-\infty}^\infty dn \left( \frac{1}{2}\rho\dot{u}_i^2 + f_{el} \right),
\]
and we introduce the total elastic energy
\[
F_{el} = \int f_{el} dV,
\]
which is a functional of the displacement and the phase field.
Thus
\begin{eqnarray*}
\frac{d{\cal W}}{dt} &=& \int\limits_{-\infty}^{\infty} dn \Bigg[ \frac{\delta F_{el}}{\delta u_i} \dot{u}_i + \frac{\delta F_{el}}{\delta \phi} \frac{\partial\phi}{\partial t} + \rho\dot{u}_i\ddot{u}_i \Bigg] \\
&=& \int\limits_{(1)}^{(2)} dn \frac{\delta F_{el}}{\delta \phi} \frac{\partial\phi}{\partial t}.
\end{eqnarray*}
The first and the last term cancel each other due to the equation of motion (\ref{phase:eq2}): pure elasticity conserves energy.
Hence, the integration interval can be restricted to a thin region around the interface, since dissipation occurs only here;
in fact, $\partial_t\phi$ decays exponentially on the scale $\xi$.
The limits of integration are inside the bulk phases 1 and 2, i.e.~a few interface widths $\xi$ away from the transition point.
This corresponds to the region of ``inner equations'', as it is typically considered for rigorous sharp interface calculations of phase field models, see e.g.~\cite{Kassner01}.
We note that upon reduction of $\xi$, the length of the integration interval becomes smaller proportionally.

We furthermore assume that the interface profile moves without shape changes, i.e. $\partial_t = -v_n \partial_n$.
Then we have
\begin{equation} \label{phase:eq3}
\frac{d{\cal W}}{dt} = -v_n \int\limits_{(1)}^{(2)} dn \frac{\delta F_{el}}{\delta \phi} \frac{\partial \phi}{\partial n}.
\end{equation}
It gives explicitly
\begin{eqnarray*}
\frac{d{\cal W}}{dt} &=& -v_n \int\limits_{(1)}^{(2)} dn \left[ \frac{\partial\mu}{\partial n} \epsilon_{ik}^2 + \frac{1}{2} \frac{\partial\lambda}{\partial n}\epsilon_{ll}^2 \right] \\
&=& -v_n \int\limits_{(1)}^{(2)} dn \frac{d}{dn} \left( \mu \epsilon_{ik}^2 + \frac{1}{2}\lambda \epsilon_{ll}^2 \right) \\
&+&  2v_n \int\limits_{(1)}^{(2)} dn \left( \mu \epsilon_{ik}\frac{\partial \epsilon_{ik}}{\partial n} + \frac{\lambda}{2}\epsilon_{ll}\frac{\partial \epsilon_{kk}}{\partial n} \right) \\
&=& -v_n \left[ \mu\epsilon_{ik}^2 + \frac{\lambda}{2}\epsilon_{ll}^2 \right]_{(1)}^{(2)} \\
&+&  v_n \int\limits_{(1)}^{(2)} dn \left( \sigma_{nn}\frac{\partial \epsilon_{nn}}{\partial n} + 2\sigma_{n\tau} \frac{\partial \epsilon_{n\tau}}{\partial n} + \sigma_{\tau\tau} \frac{\partial \epsilon_{\tau\tau}}{\partial n} \right)
\end{eqnarray*}
with the local stress $\sigma_{ij}=2\mu(\phi)\epsilon_{ij}+\lambda(\phi)\epsilon_{ll}\delta_{ij}/2$.
As we have seen above, $\epsilon_{\tau\tau}$ is continuous across the interface in the sharp interface limit, and therefore $\partial_n\epsilon_{\tau\tau}$ is finite (it only has a kink at the interface for $\xi\to 0$).
In the sharp interface limit, the integration interval becomes infinitely small, and therefore the last term in the integral, $\sigma_{\tau\tau}\partial_n\epsilon_{\tau\tau}$, does not contribute since it remains finite.
The other terms behave differently:
The stress components $\sigma_{nn}$ and $\sigma_{n\tau}$ are even continuous due to the boundary conditions.
Since they vary only smoothly on the integration interval, they can be taken out of the integrals in the sharp interface limit.
However, $\epsilon_{nn}$ and $\epsilon_{n\tau}$ are already discontinuous, and therefore, their normal derivatives contain delta function spikes at the interface.
Thus, integration gives in the limit $\xi\to 0$ e.g.~for the normal stress contribution
\[
\int\limits_{(1)}^{(2)} \sigma_{nn}\frac{\partial \epsilon_{nn}}{\partial n} dn = \sigma_{nn}\int\limits_{(1)}^{(2)}\frac{\partial\epsilon_{nn}}{\partial n} dn = \sigma_{nn}(\epsilon_{nn}^{(2)}-\epsilon_{nn}^{(1)}).
\]
Hence, we finally obtain
\begin{eqnarray}
\frac{d{\cal W}}{dt} &=& \frac{-v_n}{2} \left[ -\sigma_{nn}\epsilon_{nn} - 2\sigma_{n\tau}\epsilon_{n\tau} + \sigma_{\tau\tau} \epsilon_{\tau\tau} \right]_{(1)}^{(2)} \nonumber \\
&=& -\frac{v_n}{\Omega} (\mu^{(2)}_{el} - \mu^{(1)}_{el}) \label{phase:eq4}
\end{eqnarray}
with the chemical potential (\ref{chempot}).

On the other hand, we know that a solution of the phase field equations for an almost straight static interface is $\phi(n)=(1-\tanh(n/\xi))/2$.
Here, the value $\phi=1$ corresponds to the phase 1.
In the spirit of a rigorous sharp interface analysis, where driving force terms behave as perturbations, this is replaced by $\phi(n,t)=(1-\tanh[(n-v_n t)/\xi])/2$ if the interface starts to move, here due to elastic forces.
We insert the equation of motion (\ref{phase:eq1}) into the dissipation rate (\ref{phase:eq3}) and note that the double well potential and the surface energy do not contribute to the energy dissipation for a flat interface.
Assuming steady state motion, we obtain
\begin{equation}
\frac{d{\cal W}}{dt} = -\frac{3\gamma\xi v_n^2}{D} \int\limits_{-\infty}^\infty \left( \frac{\partial\phi}{\partial n} \right)^2 dn.
\end{equation}
Using the above phase field profile, this gives
\begin{equation}
\frac{d{\cal W}}{dt} = -\frac{3\gamma v_n^2}{4 \xi D} \int\limits_{-\infty}^\infty (1-\tanh^2\frac{n}{\xi})^2 dn = -\frac{v_n^2\gamma}{D}.
\end{equation}
Comparison with (\ref{phase:eq4}) hence gives the sharp interface limit
\begin{equation} \label{stripvnAppendix}
v_n= -\frac{D}{\gamma\Omega} (\mu^{(1)}_{el} - \mu^{(2)}_{el}).
\end{equation}
An additional curvature contribution due to surface energy then leads to Eq.~(\ref{normalvel}).

\begin{figure}
\begin{center}
\epsfig{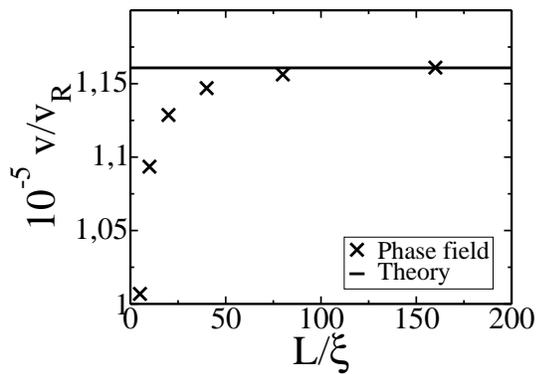}
\caption{Interface velocity $v/v_R$ for the planar front scenario as depicted in Fig.~\ref{fig2}, obtained from phase field simulations. The kinetic coefficient $D$ is small and thus the velocity remains far below the Rayleigh speed. Hence, the comparison to the quasistatic prediction Eq.~(\ref{stripvnAppendix}) leads to a good agreement if the separation of the scales $L/\xi$ is improved. In particular, the interface moves such that the soft phase extends.}
\label{PhasefieldStripvelFig}
\end{center}
\end{figure}

We also checked this scenario numerically.
For low kinetic coefficients $D$, the propagation is slow, and the dynamical code reproduces static elasticity.
Notice that the sign inversion for the normal and shear terms in the chemical potential leads to a growth of the softer phase.
We see the same behavior in the phase field simulations, and the front velocity is plotted in Fig.~\ref{PhasefieldStripvelFig}.
With increasing separation of the system size $L$ in comparison to the interface width $\xi$, the interface velocity approaches the theoretical prediction (\ref{stripvnAppendix}).

%%%%%%%%%%%%%%%%%%%%%%%%%%%%%%%%%%%%%%%%%%%%%%%%%%%%%%%%%%%%%%%%%%%%%
\section{Implementation of the Phase Field Model}
\label{ImplementationAppendix}

In this section, we explain in more detail the numerical discretization procedure which is designed to obtain a stable numerical algorithm for the elastic problem with moving boundaries.
For simplicity, we use explicit schemes for both the phase field and the elastic equations of motion.
The dissipative phase field dynamics is rather robust and therefore we do not explain the procedure here.
In contrast, the elastic equations of motion conserve energy, and tiny numerical errors can therefore easily destroy the solution.
We point out that energy conservation follows from the continuous time translation symmetry which is violated in any numerical discretization approach.
Therefore, at least fluctuations in energy are natural, but it has to be assured that the average energy does not change in time.
We experienced that naive discretization procedures can lead to long time instabilities.
The generic approach which we present here is symmetric in time and does not suffer from this problem.
It is not specifically related to the phase field description and can easily be extended to three dimensional systems or spatially varying mass densities.

\begin{figure}
\begin{center}
\epsfig{file=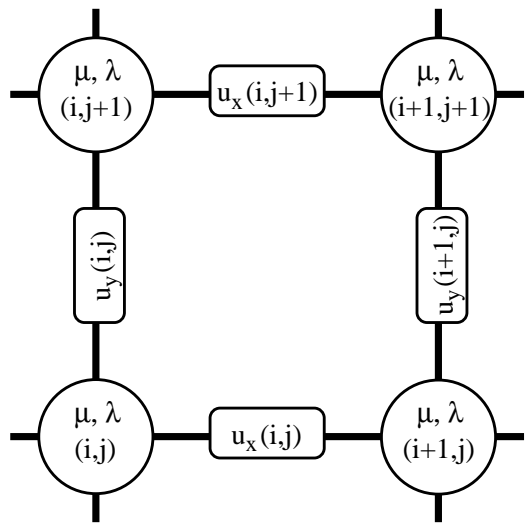, width=7cm}
\caption{The staggered grid: Shear modulus $\mu$ and Lam\'e coefficient $\lambda$ are defined on the nodes (circles), the displacements $u_i$ on the connecting lines. Thus we have three different lattices which are shifted by $\Delta x/2$.}
\label{fig3}
\end{center}
\end{figure}

We do not discuss boundary conditions and concentrate on bulk properties here.
The equation of motions can be obtained from variational principles, as was already shown in the preceding part of the manuscript.
The elastodynamic evolution Eq.~(\ref{phase:eq2}) follows from the action Eq.~(\ref{action})
\[
\frac{\delta S}{\delta u_i} = 0.
\]
We elaborate the contributions from the kinetic and the potential energy separately:
\[
S_T := \int\int \frac{1}{2}\rho \dot{u}_i^2\, dV\,dt, \qquad
S_U := -\int\int \frac{1}{2} \sigma_{ij}\epsilon_{ij}\,dV\,dt,
\]
and obtain for the potential part
\begin{eqnarray*}
S_U &=& -\frac{1}{2}\int\int \big[(2\mu+\lambda)(\epsilon_{xx}^2+\epsilon_{yy}^2) + 2\lambda \epsilon_{xx}\epsilon_{yy} \\
&& + 4\mu \epsilon_{xy}^2\big] dV\,dt.
\end{eqnarray*}
We use a staggered grid, i.e.~the mass density and the elastic constants are defined on the grid points, displacements between them (see Fig.~\ref{fig3}) \cite{Virieux86}.
In our case, the spatial (and temporal) values of the elastic coefficients $\mu, \lambda$ are related to the phase field.
Similar to the derivation above, we keep the phase field fixed (and thus the elastic coefficients) during the variation with respect to the elastic displacements.
We use the notation $u_k^{(n)}(i,j)$, where $i, j$ are the spatial and $n$ is the time index;
in the phase field formulation, no explicit distinction between the different phases has to be made, and therefore the upper index cannot be confused with previous notations.
We assume the grid spacing $\Delta x$ to be the same in both spatial directions.

The central idea for derivation of the discrete equations of motion is the discretization of the action (obeying symmetry in space and time) and to perform discrete variations with respect to each degree of freedom $u_x^{(n)}(i,j)$ and $u_y^{(n)}(i,j)$.
We study the potential contribution to $S$ first:
\begin{eqnarray*}
S_U &\rightarrow& -\frac{1}{2}(\Delta x)^2\Delta t \\
&& \sum_{n}\sum_{i,j} \Big[ \underbrace{(2\mu+\lambda)(\epsilon_{xx}^2+\epsilon_{yy}^2) + 2\lambda \epsilon_{xx}\epsilon_{yy}}_{\mbox{\tiny on grid points}} \\
&& + \underbrace{4\mu \epsilon_{xy}^2}_{\mbox{\tiny in square center}} \Big].
\end{eqnarray*}
We express the first part on the grid points, and therefore replace the elastic coefficients as follows:
\[
\mu \rightarrow \mu(i,j),\qquad \lambda \rightarrow \lambda(i,j).
\]
Strains also have to be evaluated on the nodal points:
\[
\epsilon_{xx} \rightarrow \epsilon_{xx}^{(n)}(i,j) = \frac{u_x^{(n)}(i,j)-u_x^{(n)}(i-1,j)}{\Delta x},
\]
\[
\epsilon_{yy} \rightarrow \epsilon_{yy}^{(n)}(i,j) = \frac{u_y^{(n)}(i,j)-u_y^{(n)}(i,j-1)}{\Delta x}.
\]
The second part is expressed in the center of the squares, i.~e.:
\begin{eqnarray*}
\mu &\rightarrow&  \mu(i+1/2, j+1/2) \\
=&& \!\!\!\!\!\! \frac{1}{4}\left( \mu(i,j)+\mu(i+1,j)+\mu(i,j+1)+\mu(i,j+1) \right), \\
\epsilon_{xy} &\rightarrow& \epsilon_{xy}^{(n)}(i+1/2, j+1/2) \\
&=& \Big[ u_x^{(n)}(i,j+1) - u_x^{(n)}(i,j) + u_y^{(n)}(i+1,j) \\
&& - u_y^{(n)}(i,j) \Big] /(2\Delta x).
\end{eqnarray*}
We illustrate the discrete variation with respect to $u_x^{(n)}(i,j)$,
\begin{eqnarray*}
&& \frac{\partial S_U}{\partial u_x^{(n)}(i,j)} = -\Delta t \Delta x \Big\{\left[ 2\mu(i,j)+\lambda(i,j)\right] \epsilon_{xx}^{(n)}(i,j) \\
&& - \left[ 2\mu(i+1,j)+\lambda(i+1,j)\right] \epsilon_{xx}^{(n)}(i+1,j) \\
&& + \lambda(i,j) \epsilon_{yy}^{(n)}(i,j) - \lambda(i+1,j) \epsilon_{yy}^{(n)}(i+1,j) \\
&& - 2\mu(i+1/2,j+1/2)\epsilon_{xy}^{(n)}(i+1/2,j+1/2) \\
&& + 2\mu(i+1/2,j-1/2)\epsilon_{xy}^{(n)}(i+1/2,j-1/2) \Big\}.
\end{eqnarray*}
For the kinetic contribution, we proceed in a similar way.
Here, the terms are defined between the lattice points:
\[
S_T \rightarrow \frac{1}{2} (\Delta x)^2\Delta t \sum_n\sum_{i,j} (\underbrace{\rho\dot{u}_x^2}_{\mathrm{at}\; u_x}+\underbrace{\rho\dot{u}_y^2}_{\mathrm{at}\; u_y}).
\]
Discretization of the first term defines the displacement rate $v_x^{(n+1/2)}(i,j)$ at intermediate timesteps
\[
\dot{u}_x \rightarrow v_x^{(n+1/2)}(i,j) := \frac{u_x^{(n+1)}(i,j) - u_x^{(n)}(i,j)}{\Delta t},
\]
and similarly for the second term
\[
\dot{u}_y \rightarrow v_y^{(n+1/2)}(i,j) := \frac{u_y^{(n+1)}(i,j) - u_y^{(n)}(i,j)}{\Delta t}.
\]
Variation of the kinetic contribution to the discrete action therefore gives
\begin{eqnarray*}
&& \!\!\!\!\!\! \frac{\partial S_T}{\partial u_x^{(n)}(i,j)} = - (\Delta x)^2 \rho \Big[ v_x^{(n+1/2)}(i,j) - v_x^{(n-1/2)}(i,j) \Big] \\
&=& -(\Delta x)^2 \rho \Delta t \frac{u_x^{(n+1)}(i,j)-2u_x^{(n)}(i,j) + u_x^{(n-1)}(i,j)}{(\Delta t)^2}.
\end{eqnarray*}
Notice that this expression is invariant against time inversion.
Vanishing total variation of $S=S_U+S_T$ with respect to $u_x^{(n)}(i,j)$ leads to the desired explicit evolution equation.

The same procedure has to be performed for $u_y$.

We performed various tests to check the code, among them the verification of the sound speeds.
The theoretical expressions for the dilatational and shear wave speed, $v_d=[(\lambda+2\mu)/\rho]^{1/2}$ and $v_s=(\mu/\rho)^{1/2}$ were obtained with high accuracy.
Also, we checked the transmission and reflection coefficients of both wave types at stationary interfaces.
Here, we froze the dynamics of the phase fields and let shock waves hit the straight interface between different solid phases.
The impedance of each phase is defined as $Z^{(\alpha)}=\rho v^{(\alpha)}$ with the relevant sound speed $v^{(\alpha)}$ for the considered wave type in each phase $\alpha$.
The reflection coefficient $R$ is defined as the ratio of the amplitudes of the reflected and the incoming wave and is given by
\begin{equation}
R = \frac{Z^{(2)} - Z^{(1)}}{Z^{(2)} + Z^{(1)}},
\end{equation}
and the transmission coefficient is similarly $T=1+R$.
Both values are reproduced by the numerical simulations for a phase boundary between two solids.

%%%%%%%%%%%%%%%%%%%%%%%%%%%%%%%%%%%%%%%%%%%%%%%%%%%%%%%%%%%%%%%%%%%%%
\begin{acknowledgments}
This work has been supported by the Deutsche Forschungsgemeinschaft under grant No.~SPP 1120 and the German-Israeli-Foundation.
\end{acknowledgments}

%%%%%%%%%%%%%%%%%%%%%%%%%%%%%%%%%%%%%%%%%%%%%%%%%%%%%%%%%%%%%%%%%%%%%

\end{document}